\documentclass[english]{revtex4}
\usepackage[T1]{fontenc}
\usepackage[latin9]{inputenc}
\usepackage{amsmath}
\usepackage{amssymb}

\makeatletter
\@ifundefined{textcolor}{}
{%
 \definecolor{BLACK}{gray}{0}
 \definecolor{WHITE}{gray}{1}
 \definecolor{RED}{rgb}{1,0,0}
 \definecolor{GREEN}{rgb}{0,1,0}
 \definecolor{BLUE}{rgb}{0,0,1}
 \definecolor{CYAN}{cmyk}{1,0,0,0}
 \definecolor{MAGENTA}{cmyk}{0,1,0,0}
 \definecolor{YELLOW}{cmyk}{0,0,1,0}
}

\makeatother

\usepackage{babel}
\begin{document}
\title{Local Conformal Instability and Local Non-Collapsing in the Ricci
flow of Quantum Spacetime}
\author{M.J.Luo}
\address{Department of Physics, Jiangsu University, Zhenjiang 212013, People's
Republic of China}
\email{mjluo@ujs.edu.cn}

\begin{abstract}
It is known that the conformal instability or bottomless problem rises
in the path integral method in quantizing the general relativity.
Does quantum spacetime itself really suffer from such conformal instability?
If so, does the conformal instability cause the collapse of local
spacetime region or even collapse the whole spacetime? The problems
are studied in the framework of the Quantum Spacetime Reference Frame
(QSRF) and induced spacetime Ricci flow. We find that if the lowest
eigenvalue of an operator, associated with the F-functional in a local
compact (closed and bounded) region, is positive, the local region
is conformally unstable and will tend to volume-shrinking and curvature-pinching
along the Ricci flow-time t; if the eigenvalue is negative or zero,
the local region is conformally stable up to a trivial rescaling.
However, the local non-collapsing theorem in the Ricci flow proved
by Perelman ensures that the instability will not cause the local
compact spacetime region collapse into nothing. The total effective
action is also proved positive defined and bounded from below keeping
the whole spacetime conformally stable, which can be considered as
a generalization of the classical positive mass theorem of gravitation
to the quantum level.
\end{abstract}
\maketitle

\section{Introduction}

Through a standard Wick rotation, the resulting Euclidean Einstein-Hilbert
action 
\begin{equation}
S_{E}=-\frac{1}{16\pi G}\int d^{4}x\sqrt{g}R=-\frac{1}{16\pi G}\int d^{4}x\sqrt{\tilde{g}}\left[\Omega^{2}\tilde{R}+6(\nabla\Omega)^{2}\right]\label{eq:EH+conformal}
\end{equation}
of General Relativity suffers a well known ``bottomless'' problem
\citep{Gibbons:1978ac,Hawking:1978jz}, in which $\Omega$ is a conformal
factor 
\begin{equation}
g_{\mu\nu}(x)=\Omega^{2}(x)\tilde{g}_{\mu\nu}(x).
\end{equation}
The problem is associated with the conformal instability of the theory,
a naive observation is that the kinetic term $(\nabla\Omega)^{2}$
of the conformal factor of gravity has a ``wrong sign'', making
the Euclidean action potentially become arbitrarily negative or unbounded
from below. For this reason, the functional integral quantization
of the corresponding theory
\begin{equation}
Z=\int[\mathcal{D}g_{\mu\nu}]e^{-S_{E}}\label{eq:partition}
\end{equation}
is divergent and ill defined. This divergence is not related to the
ultraviolet divergences of quantum gravity, it must be taken care
of first before one can renormalize the theory.

Various possible solutions for dealing with the conformal instability
problem have been proposed. For example, it was suggested that the
integration of the conformal factor should be performed by distorting
the integration contour in the complex plane to avoid the unboundness
action \citep{Gibbons:1978ac,Hawking:1978jz,Halliwell:1989vu}. However,
the contour deformations in functional integral would lead to complex
non-perturbative contributions to the calculations \citep{SILVESTROV1990525,David:1990ge}.
Another approach has been proposed that the theory should be formulated
in terms of dynamics and physical (transverse-traceless) degrees of
freedom by taking account of certain Jacobian factors in the functional
integral measure \citep{Biran:1983ip,Mazur:1989by,Dasgupta:2011vu},
leading to a non-standard Wick rotation \citep{Schleich:1987fm,1989Curved}
of the conformal factor. In the approach, the ``wrong sign'' of
the conformal factor can flip to be ``right'' depending on specific
range of value of an undetermined constant in the supermetric. Furthermore,
the approach works only at the linearized level, and beyond the linearized
level the Euclidean action is rather complicated, and it is not obvious
whether the ``wrong sign'' will also flip beyond the linearized
level.

An alternative possibility is that the unboundedness of the Euclidean
action may be not necessarily a real problem in defining its quantum
theory. On the one hand, there are a number of bottomless theories
at naive classical level which are well defined and having good properties
at the quantum level. For example, electron in an attractive Coulomb
well (e.g. the hydrogen atom) has a similar instability before the
quantum mechanics is discovered, since the action is also unbounded
from below which should cause the extranuclear electron falling into
the infinitely deep Coulomb well and hence the atom finally collapses.
But now it is known that the quantum mechanical treatment ultimately
evades such instability problem: the energy eigenvalue is bounded
from below and the electron has almost vanishing wavefunction and
probability amplitude to fall into the infinitely deep well. On the
other hand, similar type of conformal instability in gravitational
system may be necessary and even crucial in understanding an inflationary
universe at early epoch. For example, the conformon inflation \citep{Kallosh:2013oma}
is caused by such instability of the conformal factor: the conformon
fields. 

Without a consistent quantum theory of spacetime and gravity, the
conformal instability problem may not be cleared up and resolved.
Recently, a possible framework of quantum spacetime is proposed, i.e.
Quantum Spacetime Reference Frame (QSRF) \citep{Luo2014The,Luo2015Dark,Luo:2015pca,Luo:2019iby,Luo:2021zpi,Luo:2021wdh,Luo:2022statistics}
based on a non-linear sigma model (NLSM). The fields of the NLSM map
the coordinates of $d=4-\epsilon$ fiducial lab spacetime to a $D=4$
relativistic quantum scalar fields playing the role of frames of reference,
and it is found that the RG flow of QSRF formally generalizes the
Ricci flow in mathematics. The aim of the paper is to review the framework
and consider the conformal instability problem on the basis of the
QSRF and its induced spacetime Ricci flow.

The Ricci flow was initially introduced in 1980s by Friedan in $d=2+\epsilon$
NLSM \citep{Friedan1980,friedan1980nonlinear} and independently by
Hamilton in mathematics \citep{Hamilton1982Three,hamilton1986four}.
The main motivation of the Ricci flow from the mathematical point
of view is to classify 3-manifolds, a specific goal is to prove the
Poincare conjecture. Hamilton used it as a useful tool to gradually
deform a manifold into a more and more ``simple and good'' manifold
whose topology can be readily recognized for some simple cases. A
general realization of the program is achieved by Perelman at around
2003 \citep{perelman2002entropy,perelman2003ricci,perelman307245finite},
who introduced several monotonic functionals (the F-functional and
the W-functional) to prove the local non-collapsing theorem for finite
flow-time local singularities which may be developed in general initial
manifolds. By using the local non-collapsing theorem, he ruled out
the cigar-type of singularity during the Ricci flow, which removes
a main stumbling block of Hamilton's program to the Poincare conjecture.
Some basic theorems can be generalized to the $D=4$ Riemannian manifolds
(compact case \citep{hamilton1986four,2016Ricci}, and noncompact
\citep{1989Deforming,1989Ricci,2005Uniqueness}). 

The Ricci flow is a very powerful tool to study geometry. Perelman's
breakthrough is his discovery of the variational structure that the
Ricci flow can be seen as a gradient flow of some monotonic functionals,
it is such variational structure paves a way to study the stability
problem of a geometry. The main goal of the paper is to study the
conformal instability problem by using the monotonic functionals of
the Ricci flow of spacetime. 

Although the standard mathematical results of the Ricci flow were
oriented to prove the conjecture of 3-space, they were not generalized
directly for Lorentzian spacetime. The framework of QSRF provides
us a possible physical foundation and approach to generalize the Ricci
flow to a Riemannian or Lorentzian 4-spacetime, when the relativistic
quantum frame fields are considered measuring rods and clocks of the
spacetime. The Ricci flow of the spacetime is then considered as a
smearing and coarse-graining process induced by 2nd order central
moment quantum fluctuations of the spacetime frame fields. Furthermore,
monotonic functionals of Perelman\textquoteright s type are also deduced
from the diffeomorphism anomaly in the functional integral quantization
of QSRF. So we could consider QSRF may lay a possible physical foundation
to the spacetime Ricci flow and related monotonic functional of Perelman
for physical spacetime. In this sense, at the physical rigorous level,
the Ricci flow and Perelman's monotonic functionals can be formally
generalized to the Lorentzian spacetime. Particularly, the physical
important spacetime configurations are those Ricci solitons as the
flow limit. The Ricci soliton equation weakly cares about the spacetime
signature, and the signature of its solution depends on the boundary
condition. The Ricci flow of a Lorentzian manifolds is also an important
direction in geometry (see e.g. \citep{2010arXiv1007.3397B,2011arXiv1106.2924B,2014arXiv1403.4400B}).
We have noticed that even without a clear and generally accepted physical
foundation, there already have formal applications of the Ricci flow
to the physical spacetime (see e.g. \citep{Gutperle:2002ki,2005Mass,2006Ricci,2006A,Kholodenko:2007ep,Husain:2008rg,Ruchin:2013azz,Cartas-Fuentevilla:2017cvt,Cartas-Fuentevilla:2018rez}),
although they may lead to some conceptual confusions in interpreting
their results. We also notice that various relativistic generalization
and non-Riemannian modifications of Perelman's works have been considered
in gravity and information theories (see e.g. \citep{2020AnPhy.42368333B,2020EPJC...80..639V,2019arXiv190703541B}
and references therein), which is important for a consistent theory
of the spacetime Ricci flow. 

The structure of the paper is as follows. In the section II, we briefly
review the framework of QSRF and its induced Ricci flow of spacetime,
in which monotonic functionals of Perelman's type are deduced from
the diffeomorphism anomaly of QSRF, and the effective gravity theory
of QSRF is also discussed. In the section III, by using the monotonic
F-functional of the theory which generalizes Perelman's F-functional,
the conformal instability problem can be reconsidered within the framework.
In the section IV, by using the non-local-collapsing theorem and related
W-functional developed by Perelman, the final destiny of the local
conformal instability of quantum spacetime can be studied. Finally,
we discuss and conclude the paper in section V.

\section{A Theory of Quantum Spacetime}

To make the paper self-contained, in this section, we briefly review
the framework of Quantum Spacetime Reference Frame (QSRF) as a possible
theory of quantum spacetime for a preliminary, which have certain
overlaps with the previous works \citep{Luo2014The,Luo2015Dark,Luo:2015pca,Luo:2019iby,Luo:2021zpi,Luo:2021wdh,Luo:2022statistics}.
The section provides a framework and methodology to the conformal
instability problem. We hope this section provides some general basis
and background of the Ricci flow and related effective gravity theory
to the readers. If the readers have been familiar with related literature,
they could skip the section and directly go to the section III and
IV.

\subsection{Quantum Spacetime Reference Frame (QSRF)}

The quantum spacetime reference frame starts from the idea that spacetime
is nothing but an ideal and standard reference that a moving body
is relative to. In fact ``ideal'' can not be realized in rigor at
the quantum level, all things are quantum fluctuating. A quantum standard
reference can be measured (relative to a fiducial lab) by quantum
rods and quantum clocks which are also subject to quantum fluctuations.
Surprisingly, gravitation is nothing but emergent phenomenon coming
from relative measurement of the under-studied moving body being reference
to the quantum standard reference, i.e. quantum reference frame. If
the classical equivalence principle is generalized to the quantum
level, then some universal (classical or quantum) properties (e.g.
the universal acceleration of a free falling body which is independent
to its mass, the Hubble redshift (universal recession velocity) and
even the broadening (universal acceleration) of spectral lines are
independent to theirs energies) are not merely the (classical or quantum)
properties of matter, they are actually the (classical or quantum)
properties of the spacetime itself. Thus in this framework, the most
fundamental thing is not individual quantum state but the relations
between two quantum states, i.e. the under-studied quantum body $|\psi\rangle$
and the quantum reference frame system $|X\rangle$, described by
an entangled state $|\Psi\rangle=\sum_{ij}|\psi\rangle_{i}\otimes|X\rangle_{j}$
in the whole Hilbert space $\mathcal{H}_{\psi}\otimes\mathcal{H}_{X}$.

If the quantum spacetime reference frame $|X_{\mu}\rangle$ ($\mu=0,1,2,...D-1$)
itself is considered as the under-studied quantum body, then we have
a quantum theory of the quantum spacetime. In this case the reference
system could be the fiducial lab spacetime $|x_{a}\rangle$, ($a=0,1,2,...d-1$).
The entangled state is constructed by a one-to-one correspondence
between two states, i.e. $|x_{a}\rangle\rightarrow|X_{\mu}\rangle$,
which is nothing but a general non-linear differentiable mapping $X:\mathbb{R}^{d}\rightarrow M^{D}$.
From the geometric point of view, the entangled state $\sum_{ij}|X\rangle_{i}\otimes|x\rangle_{j}$
or $X(x)$ is a mapping from a local coordinate flat patch $x\in\mathbb{R}^{d}$
to a Riemannian or Lorentzian manifolds $X\in M^{D}$ (the signature
depends on the boundary condition). From the physics point of view,
the general non-linear mapping can be realized by a kind of field
theory, the non-linear sigma model (NLSM)
\begin{equation}
S[X]=\frac{1}{2}\lambda\int d^{d}xg_{\mu\nu}\frac{\partial X^{\mu}}{\partial x_{a}}\frac{\partial X^{\nu}}{\partial x_{a}},
\end{equation}
where $\lambda$ is a constant with dimension of energy density $[L^{-d}]$
taking the value of the critical density (\ref{eq:critical density})
of the universe. $x_{a}$ having dimension of length $[L]$ is the
base space of NLSM. It will be interpreted as the lab wall and clock
frame as the starting reference, which is fiducial and classical (infinite
precise and free from quantum fluctuations). Frame field $X_{\mu}$
also having dimension of length is the target space of NLSM interpreted
as physical rods and clocks, i.e. the reference frame fields measuring
the quantum spacetime coordinate. $X_{\mu}$ are the coordinates of
a general Riemannian or Lorentzian manifolds $M^{D}$ with curved
metric $g_{\mu\nu}$, called the target space in NLSM's terminology.
We will work with the real-defined coordinates for the target spacetime,
and the Wick rotated case has been generally included into the general
coordinates transformation of the time component. The frame fields
will be promoted to quantum fields, in the language of quantum fields
theory, $X_{\mu}(x)$ or $X^{\mu}(x)=g^{\mu\nu}X_{\nu}(x)$ are the
real scalar frame fields. $D\equiv4$ is the least number of the frame
fields capable to measure the coordinates of a spacetime event. $d$
is the dimensions of the fiducial lab spacetime, it could take $d=4-\epsilon$,
where $0<\epsilon\ll1$ is for topological reason that the homotopic
group $\pi_{d}(M^{D})$ should be trivial so that the mapping $X(x):\mathbb{R}^{d}\rightarrow M^{D}$
should be free from topological obstacles and singularities. Since
a quantum fields theory is well-formulated in an inertial frame, and
the local patch is also flat, so the fiducial lab spacetime $x_{a}$
is considered flat and rigid. Note that $d^{d}x\equiv d^{d}x\det e$
($\det e$ is the Jacobian with $|\det e|=1$) is a classical invariant
volume element no matter in Minkovskian or Euclidean, since if one
replaces the Euclidean time to the Minkowskian time $dx_{0}^{(E)}\rightarrow idx_{0}^{(M)}$,
the Jacobian changes correspondingly $\det e^{(E)}\rightarrow i\det e^{(M)}$,
so the Euclidean theory will give the same result as the Minkowskian
one at least classically, and the action is always real no matter
it is Minkowskian or Euclidean one. Without loss of generality, we
consider the base space as the Euclidean one which is better defined
when one uses the functional integral quantization method. It is well-known
that the NLSM in $d=2$ is perturbative renormalizable, and some numerical
calculations also support $d=3$ and $d=4-\epsilon<4$ are non-perturbative
renormalizable, so it is a well-defined relativistic quantum fields
theory.

The quantum frame fields model of spacetime has practical physical
interpretation, for instance, in the lab scale, it can be considered
as a multi-wire proportional chamber system constructed relative to
the wall and clock of a lab that are used to measure coordinates of
events in the lab, where the frame fields can be interpreted as the
spin-less electron signal. When the spacetime measuring scale is beyond
the lab's scale, for instance, to the cosmic scale, the quantum fluctuation
or broadening of the frame fields (e.g. by using spin-less light signal)
become unignorable, and curved metric $g_{\mu\nu}$ measured by the
comparison between the frame fields and fiducial coordinates become
non-trivial and important, since the quantum fluctuation $X_{\mu}=\langle e_{\mu}^{a}\rangle x_{a}+\delta X_{\mu}$
has nontrivial physical consequence. The lowest 2nd order central
moment (variance) modifies the quadratic form distance of Riemannian
or Lorentzian geometry and hence gives correction to the quantum expectation
of the metric
\begin{equation}
\langle g_{\mu\nu}\rangle=\left\langle \frac{\partial X_{\mu}}{\partial x_{a}}\frac{\partial X_{\nu}}{\partial x_{a}}\right\rangle =\left\langle \frac{\partial X_{\mu}}{\partial x_{a}}\right\rangle \left\langle \frac{\partial X_{\nu}}{\partial x_{a}}\right\rangle +\frac{\partial^{2}}{\partial x_{a}^{2}}\left\langle \delta X_{\mu}\delta X_{\nu}\right\rangle =g_{\mu\nu}^{(1)}(X)+\delta g_{\mu\nu}^{(2)}(X).\label{eq:g(1)+g(2)}
\end{equation}
The 2nd order moment quantum fluctuation \citep{percacci2009asymptotic,codello2009fixed}
\begin{equation}
\delta g_{\mu\nu}^{(2)}(X)=\frac{R_{\mu\nu}^{(1)}(X)}{32\pi^{2}\lambda}\delta k^{d-2}
\end{equation}
deforms the metric, where $R_{\mu\nu}^{(1)}$ is the Ricci curvature
given by $g_{\mu\nu}^{(1)}$ and $k^{d-2}$ is the cutoff energy scale
of the Fourier component of the frame fields. For $d=4-\epsilon$,
the validity of the perturbation calculation $R^{(1)}\delta k^{2}\ll\lambda$
is the validity of the Gaussian approximation. It will be shown later
that $\lambda$ is nothing but the critical density $\rho_{c}$ of
the universe, $\lambda\sim O(H_{0}^{2}/G)$, $H_{0}$ the current
Hubble's parameter, $G$ the Newton's constant. Thus for our concern
of pure gravity in which matter is ignored, the condition $R^{(1)}\delta k^{2}\ll\lambda$
is equivalent to $\delta k^{2}\ll1/G$ which is reliable except for
some local singularities may develop when the Gaussian approximation
is failed.

\subsection{The RG flow of QSRF as the Ricci flow of Spacetime}

The deformation of the metric at the Gaussian approximation is driven
by the 2nd moment quantum fluctuation and hence the Ricci curvature,
which gives rise to the Ricci flow equation \citep{Hamilton1982Three,hamilton1986four}
\begin{equation}
\frac{\partial g_{\mu\nu}}{\partial t}=-2R_{\mu\nu},
\end{equation}
where the flow-time interval
\begin{equation}
\delta t=-\frac{1}{64\pi^{2}\lambda}\delta k^{d-2}
\end{equation}
has the dimension of distance squared $[L^{2}]$ for any dimension
of the base space $d$, in our relativistic quantum frame fields setting,
we have taken $d=4-\epsilon$. 

For the Ricci curvature is non-linear for the metric, the Ricci flow
equation is a non-linear version of the heat equation for the metric,
the metric flowing along $t$ introduces an averaging or coarse-graining
process to the intrinsic non-linear quantum spacetime which is highly
non-trivial. In general, if the flow is free from local singularities
there exists long flow-time solution in $t\in(-\infty,0)$, which
is often called ancient solution in mathematical literature. This
range of the t-parameter corresponds to $k\in(0,\infty)$, that is
from $t=-\infty$, i.e. the short distance (high energy) UV scale
$k=\infty$ forwardly to $t=0$ i.e. the long distance (low energy)
IR scale $k=0$. The metric at certain scale $t$ is given by being
averaged out the shorter distance details which produces an effective
correction to the metric at that scale. So along t, the manifolds
loss its information in shorter distance, thus the flow is irreversible,
i.e. generally having no backwards solution, which is the underlying
reason for the entropy of a spacetime. 

As it is shown in (\ref{eq:g(1)+g(2)}), the 2nd order moment fluctuation
modifies the local (quadratic) distance of the spacetime, so the flow
is non-isometry. The non-isometry is not important for its topology,
so along t, the flow preserves the topology of the spacetime but its
local metric, shape and size (volume) change. There also exists a
very special solution of the Ricci flow called Ricci soliton, which
only changes the local volume while keeps its local shape. The Ricci
soliton, and its generalized version, the Gradient Ricci Soliton,
as the flow limits (of the generalized Ricci-DeTurck flow), are the
generalization of the notion of fixed point in the sense of RG flow.
The solution of the Ricci soliton as the fixed point solution of the
Ricci flow can be either Riemannian or Lorentzian manifolds, whose
signature is determined by the boundary condition of the solution. 

Although the Ricci flow is not strongly parabolic, DeTurck provides
a simple way to prove the short-flow-time existence of the Ricci flow
by slightly modifying it, named Ricci-DeTurck flow \citep{deturck1983deforming}
\begin{equation}
\frac{\partial g_{\mu\nu}}{\partial t}=-2\left(R_{\mu\nu}+\nabla_{\mu}\nabla_{\nu}f\right),\label{eq:ricci-deturk}
\end{equation}
which is equivalent to the standard Ricci flow equation up to a diffeomorphism
given by $u=e^{-f}$. From the geometric point of view, $u=e^{-f}$
introduces a positive defined density bundle over the local geometric,
so it is also called a density of manifolds. From the physics point
of view $u$ is referred to as a conformal factor up to a constant
multiple which compensates the flow of the volume (i.e. the longitudinal
degrees of freedom of gravity). From the statistic physics point of
view, $u$ can also be interpreted as a positive defined density matrix
\citep{Luo:2022statistics} of the frame fields system, $u$ is usually
normalized by the condition 
\begin{equation}
\lambda\int ud^{D}X=1,\label{eq:u-normalization}
\end{equation}
which also defines an invariant and fiducial volume (the rigid lab
frame) from the language of density manifolds. The condition, together
with the Ricci-DeTurck flow, gives the flow equation of $u$, 
\begin{equation}
\frac{\partial u}{\partial t}=\left(-\Delta+R\right)u.\label{eq:u-t-eq}
\end{equation}
Note the minus sign in front of the Laplacian, it is a backwards heat-like
equation. Naively speaking, the solution of the backwards heat flow
will not exist. But we could also note that if one let the Ricci flow
flows to certain singular IR scale $t_{*}$, (if the flow is free
from global singularity from the mapping $X(x):\mathbb{R}^{d}\rightarrow M^{D}$,
the solution is ancient, and hence $t_{*}=0$), and at $t_{*}$ one
may then choose an appropriate $u(t_{*})=u_{0}$ arbitrarily (up to
a diffeomorphism gauge) and flows it backwards in $\tau=t_{*}-t$
to obtain a solution $u(\tau)$ of the backwards equation. In this
case, $u$ satisfies the heat-like equation 
\begin{equation}
\frac{\partial u}{\partial\tau}=\left(\Delta-R\right)u,\label{eq:conjugate heat eq}
\end{equation}
which does admit a solution along $\tau$, often called the conjugate
heat equation in mathematical literature.

So far (\ref{eq:conjugate heat eq}) together with (\ref{eq:ricci-deturk})
the mathematical problem of the Ricci flow of a Riemannian/Lorentzian
manifolds is transformed to coupled equations
\begin{equation}
\begin{cases}
\frac{\partial g_{\mu\nu}}{\partial t}=-2\left(R_{\mu\nu}-\nabla_{\mu}\nabla_{\nu}f\right)\\
\frac{\partial u}{\partial\tau}=\left(\Delta-R\right)u\\
\frac{d\tau}{dt}=-1
\end{cases}
\end{equation}
and the manifolds $(M^{D},g)$ are generalized to density manifolds
$(M^{D},g,u)$ \citep{Morgan2009Manifolds,2016arXiv160208000W,Corwin2017Differential}
with the constraint (\ref{eq:u-normalization}). 

\subsection{Anomaly Induced Action of QSRF}

The quantum fluctuation and hence the Ricci flow does not preserve
the quadratic distance of a Riemannian or Lorentzian geometry. The
non-isometry of the quantum fluctuation induces a breakdown of diffeomorphism
or general coordinate transformation at the quantum level, namely
the anomaly. It will be surprise to see that the monotonic functionals
of Perelman's type and an effective action of gravity can be deduced
from the anomaly of QSRF.

Here we consider functional quantization of the pure frame fields,
the partition function is
\begin{equation}
Z(M^{D})=\int[\mathcal{D}X]\exp\left(-S[X]\right)=\int[\mathcal{D}X]\exp\left(-\frac{1}{2}\lambda\int d^{4}xg^{\mu\nu}\partial_{a}X_{\mu}\partial_{a}X_{\nu}\right),
\end{equation}
where $M^{D}$ is the target spacetime, and the base space is Euclidean.
Note that a general coordinate transformation
\begin{equation}
X_{\mu}\rightarrow\hat{X}_{\mu}=\frac{\partial\hat{X}_{\mu}}{\partial X_{\nu}}X_{\nu}=e_{\mu}^{\nu}X_{\nu}
\end{equation}
does not change the action $S[X]=S[\hat{X}]$, but the measure of
the functional integral changes
\begin{align}
\mathcal{D}\hat{X} & =\prod_{x}\prod_{\mu=0}^{D}d\hat{X}_{\mu}(x)=\prod_{x}\epsilon_{\mu\nu\rho\sigma}e_{\mu}^{0}e_{\nu}^{1}e_{\rho}^{2}e_{\sigma}^{3}dX_{0}(x)dX_{1}(x)dX_{2}(x)dX_{3}(x)\nonumber \\
 & =\prod_{x}\left|\det e(x)\right|\prod_{x}\prod_{a=0}^{D}dX_{a}(x)=\left(\prod_{x}\left|\det e(x)\right|\right)\mathcal{D}X,
\end{align}
where
\begin{equation}
\left|\det e_{\mu}^{a}\right|=\sqrt{\left|\det g_{\mu\nu}\right|}
\end{equation}
is the Jacobian of the diffeomorphism. The Jacobian is nothing but
a local relative volume element $dV(\hat{X}_{\mu})$ w.r.t. the fiducial
volume $dV(X_{a})$. Note that the normalization condition (\ref{eq:u-normalization})
also defines a fiducial volume element $ud^{4}X\equiv udV(\hat{X}_{\mu})$,
so the Jacobian is nothing but related to the volume ratio
\begin{equation}
u(\hat{X})=\frac{dV(X_{a})}{dV(\hat{X}_{\mu})}=\left|\det e_{a}^{\mu}\right|=\frac{1}{\left|\det e_{\mu}^{a}\right|}.\label{eq:volume form}
\end{equation}

Until now, the previous derivations are straightforward, the above
$u$ density followed by (\ref{eq:u-normalization}) is the first
explicit generalization from the standard 3-space to a Riemannian
or Lorentzian 4-spacetime. It is worth stressing that here the absolute
symbol of the determinant is because the density $u$ and the volume
form are kept positive defined even in the Lorentz signature. Otherwise,
for the Lorentz signature, there should introduce some extra imaginary
factor $i$ into (\ref{eq:u}) to keep the condition (\ref{eq:u-normalization}).
In fact the definition of the volume form and the manifolds density
ensure the formalism of the framework formally the same with the Perelman's
standard form even in the Lorentzian signature. The manifolds density
encodes a most important information of a Riemannian or Lorentzian
geometry, i.e. the local volume comparison. 

In this situation, if we parameterize a dimensionless solution $u$
of the conjugate heat equation (\ref{eq:conjugate heat eq}) as 
\begin{equation}
u(\hat{X})=\frac{1}{\lambda(4\pi\tau)^{D/2}}e^{-f(\hat{X})},\label{eq:u}
\end{equation}
then the partition function $Z(M^{D})$ is transformed to
\begin{align}
Z(\hat{M}^{D}) & =\int[\mathcal{D}\hat{X}]\exp\left(-S[\hat{X}]\right)=\int\left(\prod_{x}\left|\det e\right|\right)[\mathcal{D}X]\exp\left(-S[X]\right)\nonumber \\
 & =\int\left(\prod_{x}e^{f+\frac{D}{2}\log(4\pi\tau)}\right)[\mathcal{D}X]\exp\left(-S[X]\right)\nonumber \\
 & =\int[\mathcal{D}X]\exp\left\{ -S[X]+\lambda\int d^{4}x\left[f+\frac{D}{2}\log(4\pi\tau)\right]\right\} \nonumber \\
 & =\int[\mathcal{D}X]\exp\left\{ -S[X]+\lambda\int_{\hat{M}^{D}}d^{D}Xu\left[f+\frac{D}{2}\log(4\pi\tau)\right]\right\} .
\end{align}

Note that the change of the partition function 
\begin{equation}
Z(\hat{M}^{D})=e^{\lambda N(\hat{M}^{D})}Z(M^{D})\label{eq:Z->Zhat}
\end{equation}
is nothing but a pure real Shannon entropy in terms of the density
$u$
\begin{equation}
N(\hat{M}^{D})=\int_{\hat{M}^{D}}d^{D}Xu\left[f+\frac{D}{2}\log(4\pi\tau)\right]=-\int_{\hat{M}^{D}}d^{D}Xu\log u.\label{eq:shannon entropy}
\end{equation}

The classical action $S[X]$ is invariant under the general coordinates
transformation or diffeomorphism, but the quantum partition function
is no longer invariant under the general coordinates transformation
or diffeomorphism, which is called diffeomorphism anomaly, meaning
a breaking down of the diffeomorphism at the quantum level. The diffeomorphism
anomaly is purely due to the quantum fluctuation and Ricci flow of
the frame fields which do not preserve the functional integral measure
and change the spacetime volume at the quantum level. The diffeomorphism
anomaly has many profound consequences to the theory of quantum reference
frame, e.g. non-unitarity, the trace anomaly, the notion of entropy,
reversibility, and the cosmological constant \citep{Luo:2021zpi,Luo:2022statistics}.

Without loss of generality, if we simply consider the under-transformed
coordinates $X_{\mu}$ as the coordinates of the fiducial lab $x_{a}$
which can be treated as a classical parameter coordinates, in this
situation the classical action of NLSM is just a topological invariant,
i.e. half the dimension of the target spacetime
\begin{equation}
\exp\left(-S_{cl}\right)=\exp\left(-\frac{1}{2}\lambda\int d^{4}xg_{(1)}^{\mu\nu}\partial_{a}x_{\mu}\partial_{a}x_{\nu}\right)=\exp\left(-\frac{1}{2}\lambda\int d^{4}xg_{(1)}^{\mu\nu}g_{\mu\nu}^{(1)}\right)=e^{-\frac{D}{2}}.
\end{equation}
Thus the total partition function of the frame fields takes a simple
form
\begin{equation}
Z(\hat{M}^{D})=e^{\lambda N(\hat{M}^{D})-\frac{D}{2}}.\label{eq:frame-partition}
\end{equation}
A relative density 
\begin{equation}
u_{r}(X)=\frac{u}{u_{*}}
\end{equation}
can be defined by a density $u(X)$ being relative to 
\begin{equation}
u_{*}(X)=\frac{1}{\lambda(4\pi\tau)^{D/2}}\exp\left(-\frac{1}{4\tau}\left|g_{\mu\nu}X^{\mu}X^{\nu}\right|\right),
\end{equation}
in which the absolute symbol in the exponential is used to keep the
quadratic distance and hence the volume form induced from the Gaussian
integral positive even in the Lorentzian signature. $u_{*}$ corresponds
to the maximum Shannon entropy $N_{*}$
\begin{equation}
N_{*}=-\int d^{D}Xu_{*}\log u_{*}=\int d^{D}Xu_{*}\frac{D}{2}\left[1+\log(4\pi\tau)\right]=\frac{D}{2\lambda}\left[1+\log(4\pi\tau)\right],
\end{equation}
when the spacetime becomes a gradient shrinking Ricci soliton (GSRS)
satisfying
\begin{equation}
R_{\mu\nu}+\nabla_{\mu}\nabla_{\nu}f=\frac{1}{2\tau}g_{\mu\nu}.\label{eq:shrinker}
\end{equation}
By using the relative density, a relative Shannon entropy $\tilde{N}$
can be defined by
\begin{equation}
\tilde{N}(M^{D})=-\int d^{D}Xu\log u_{r}=-\int d^{D}Xu\log u+\int d^{D}Xu_{*}\log u_{*}=N-N_{*}=-\log Z_{P}\le0,\label{eq:perelman-partition}
\end{equation}
where $Z_{P}$ is nothing but the Perelman's partition function \citep{perelman2002entropy}
\begin{equation}
\log Z_{P}=\int_{M^{D}}d^{D}Xu\left(\frac{D}{2}-f\right)\ge0.
\end{equation}
In terms of the relative Shannon entropy, the total partition function
(\ref{eq:frame-partition}) of the frame fields is normalized by the
extreme value 
\begin{equation}
Z(M^{D})=\frac{e^{\lambda N-\frac{D}{2}}}{e^{\lambda N_{*}}}=e^{\lambda\tilde{N}-\frac{D}{2}}.\label{eq:relative-partition}
\end{equation}

\subsection{Effective Theory of Gravity}

Followed by different route to generalize the Ricci flow, a Ricci
flow of a Lorentzian 4-spacetime is constructed on the physical foundation
of the QSRF, which may provide us an alternative way to generalize
the 3-space Ricci flow to the 4-spacetime version, we take $D=4$
in this subsection. Together with a quantum generalization of Equivalence
Principle, an effective gravity is emerged from the QSRF. 

The relative Shannon entropy $\tilde{N}$ as the anomaly vanishes
at gradient shrinking Ricci soliton (GSRS) or IR scale, however, it
is non-zero at ordinary lab scale up to UV where the fiducial volume
of the lab is considered rigid and fixed $\lambda\int d^{4}x=1$.
The cancellation of the anomaly at the lab scale up to UV is physically
required, which leads to the counter term $\nu(M_{\tau=\infty}^{D})$
or cosmological constant. Since the $\tilde{N}$ is monotonic non-decreasing
along $t$ or non-increasing along $\tau$
\begin{equation}
\frac{\partial\tilde{N}(M^{D})}{\partial\tau}=\frac{\partial N(M^{D})}{\partial\tau}-\frac{\partial N_{*}}{\partial\tau}=\mathcal{F}(M^{D})-\frac{D}{2\lambda\tau}\equiv\tilde{\mathcal{F}}(M^{D})\le0.\label{eq:Ntilde_monontonic}
\end{equation}
The monotonicity of $\tilde{N}$ and the W-functional implies a lower
bound at UV \citep{2003math......6147N}
\begin{equation}
\nu(M_{\tau=\infty}^{D})=\lim_{\tau\rightarrow\infty}\lambda\tilde{N}(M^{D},u,\tau)=\lim_{\tau\rightarrow\infty}\lambda\mathcal{W}(M^{D},u,\tau)=\inf_{\tau}\lambda\mathcal{W}(M^{D},u,\tau)<0,
\end{equation}
where $\mathcal{W}$, the W-functional, is the Legendre transformation
of $\tilde{N}$ w.r.t. $\tau^{-1}$,
\begin{align}
\mathcal{W} & \equiv\tau\frac{\partial\tilde{N}}{\partial\tau}+\tilde{N}=\tau\tilde{\mathcal{F}}+\tilde{N}.\label{eq:W=00003DtauF+N}
\end{align}
The Ricci-DeTurck flow turns out to be a gradient flow of the monotonic
functionals. The functionals are formally identified with Perelman's,
with the monotonicities \citep{perelman2002entropy}
\begin{equation}
\frac{\partial\mathcal{F}}{\partial t}\ge\frac{2}{D}\int d^{D}Xu\left(R+\Delta f\right)^{2}\ge\frac{2}{D}\left[\int d^{D}Xu\left(R+\Delta f\right)\right]^{2}=\frac{2}{D}\mathcal{F}^{2}\ge0,\label{eq:F-monotonic}
\end{equation}
\begin{equation}
\frac{\partial\mathcal{W}}{\partial t}=2\tau\int d^{D}Xu\left|R_{\mu\nu}+\nabla_{\mu}\nabla_{\nu}f-\frac{1}{2\tau}g_{\mu\nu}\right|^{2}\ge0.\label{eq:W_monotonic}
\end{equation}
In this sense, QSRF provides us a framework to calculate and generalize
Perelman's W-functional to a Lorentzian spacetime by using the anomaly
or Shannon entropy in terms of the density $u$ given by the local
volume form of a Lorentzian geometry (\ref{eq:volume form}).

In other words, the difference between the effective actions (relative
Shannon entropies) at UV and IR is finite
\begin{equation}
\nu=\lambda(\tilde{N}_{\tau=\infty}-\tilde{N}_{\tau=0})<0.
\end{equation}

In fact $e^{\nu}<1$ (usually called the Gaussian density \citep{cao2004gaussian,cao2009recent})
is a relative volume or the reduced volume $\tilde{V}(M_{\tau=\infty}^{D})$
of the backwards limit manifolds introduced by Perelman, or the inverse
of the initial condition of the manifolds density $u_{\tau=0}^{-1}$.
A finite value of it makes an initial spacetime with unit volume from
UV flow and converge to a finite $u_{\tau=0}$, and hence the manifolds
finally converge to a finite relative volume/reduced volume instead
of shrinking to a singular point at $\tau=0$.

As an example, for a homogeneous and isotropic universe for which
the sizes of space and time (with a ``ball'' radius $a_{\tau}$)
are on an equal footing, i.e. $ds^{2}=a_{\tau}^{2}(-dx_{0}^{2}+dx_{1}^{2}+dx_{2}^{2}+dx_{3}^{2})$,
which is a Lorentzian shrinking soliton configuration. Note that the
shrinking soliton equation $R_{\mu\nu}=\frac{1}{2\tau}g_{\mu\nu}$
it satisfies and its volume form (\ref{eq:volume form}) are independent
to the signature, so it can be approximately given by a 4-ball value
$\nu(B_{\infty}^{4})\approx-0.8$ \citep{Luo:2019iby,Luo:2021zpi}.

So the partition function, which is anomaly canceled at UV and having
a fixed-volume fiducial lab, is
\begin{equation}
Z(M^{D})=e^{\lambda\tilde{N}-\frac{D}{2}-\nu}.\label{eq:partition with nu}
\end{equation}
Since $\lim_{\tau\rightarrow0}\tilde{N}(M^{D})=0$, so at small $\tau$,
$\tilde{N}(M^{D})$ can be expanded by powers of $\tau$
\begin{align}
\tilde{N}(M^{D}) & =\frac{\partial\tilde{N}}{\partial\tau}\tau+O(\tau^{2})=\tau\tilde{\mathcal{F}}+O(\tau^{2})\nonumber \\
 & =\int_{M^{D}}d^{D}Xu\left[\left(R_{\tau=0}+\left|\nabla f_{\tau=0}\right|^{2}-\frac{D}{2\tau}\right)\tau\right]+O(\tau^{2})\nonumber \\
 & =\int_{M^{D}}d^{D}XuR_{0}\tau+O(\tau^{2}),\label{eq:Ntild=00003Dtau*Ftild}
\end{align}
in which $\lambda\int d^{D}Xu\tau\left|\nabla f_{\tau\rightarrow0}\right|^{2}\approx\frac{D}{2}$
(at GSRS) has been used.

For $D=4$ and small $\tau$, the effective action of $Z(M^{4})$
can be given by
\begin{equation}
-\log Z(M^{4})=S_{eff}\approx\int_{M^{4}}d^{4}Xu\left(2\lambda-\lambda R_{0}\tau+\lambda\nu\right)\quad(\textrm{small\,\ensuremath{\tau}}).
\end{equation}
Considering $ud^{4}X$ as the invariant and fiducial volume element
$d^{4}X\sqrt{|g_{k}|}$ at certain scale $k$ and $\tau=-t=\frac{1}{64\pi^{2}\lambda}k^{2}$
(when $t_{*}=0$ i.e. the global geometry is free from singularity),
we have
\begin{equation}
S_{eff}=\int_{M^{4}}d^{4}X\sqrt{|g_{k}|}\left(2\lambda-\frac{R_{0}}{64\pi^{2}}k^{2}+\lambda\nu\right)\quad(\textrm{small\,k}).\label{eq:eff-EH}
\end{equation}

The effective action can be interpreted as a low energy effective
action of pure gravity. As the cutoff scale $k$ ranges from the lab
scale to the solar system scale ($k>0$), the action must recover
the well-tested Einstein-Hilbert (EH) action. But at the cosmic scale
($k\rightarrow0$), we know that the EH action deviates from observations
and the cosmological constant becomes important. In this picture,
as $k\rightarrow0$, the action leaving $2\lambda+\lambda\nu$ should
play the role of the standard EH action with a limit constant background
scalar curvature $R_{0}$ plus the cosmological constant, so
\begin{equation}
2\lambda+\lambda\nu=\frac{R_{0}-2\Lambda}{16\pi G}.
\end{equation}
While at UV $k\rightarrow\infty$, $\lambda\tilde{N}\rightarrow\nu$,
the action leaves only the fiducial Lagrangian $\frac{D}{2}\lambda=2\lambda$
which should be interpreted as a constant EH action without the cosmological
constant
\begin{equation}
2\lambda=\frac{R_{0}}{16\pi G}.
\end{equation}
Thus we have the cosmological term 
\begin{equation}
\lambda\nu=\frac{-2\Lambda}{16\pi G}=-\rho_{\Lambda}.
\end{equation}
The action can be rewritten as an effective EH action plus a cosmological
term
\begin{equation}
S_{eff}=\int_{M^{4}}d^{4}X\sqrt{|g_{k}|}\left(\frac{R_{k}}{16\pi G}+\lambda\nu\right)\quad(\textrm{small\,k}),\label{eq:analogy EH}
\end{equation}
where
\begin{equation}
\frac{R_{k}}{16\pi G}=2\lambda-\frac{R_{0}}{64\pi^{2}}k^{2},\label{eq:eff-R}
\end{equation}
which is nothing but the flow equation of the scalar curvature
\begin{equation}
R_{k}=\frac{R_{0}}{1+\frac{1}{4\pi}Gk^{2}},\quad\textrm{or}\quad R_{\tau}=\frac{R_{0}}{1+\frac{2}{D}R_{0}\tau}.
\end{equation}
Since at the cosmic scale $k\rightarrow0$, the effective scalar curvature
is bounded by $R_{0}$ which can be measured by the ``Hubble's constant''
$H_{0}$ at the cosmic scale,
\begin{equation}
R_{0}=D(D-1)H_{0}^{2}=12H_{0}^{2},
\end{equation}
so $\lambda$ is nothing but the critical density of the 4-spacetime
Universe
\begin{equation}
\lambda=\frac{3H_{0}^{2}}{8\pi G}=\rho_{c},\label{eq:critical density}
\end{equation}
so the cosmological constant is always of order of the critical density
with a ``dark energy'' fraction
\begin{equation}
\Omega_{\Lambda}=\frac{\rho_{\Lambda}}{\rho_{c}}=-\nu\approx0.8,
\end{equation}
which is not far from observations. The detailed discussions about
the cosmological constant problem and the observational effect in
the cosmology, especially the modification of the Distance-Redshift
relation leading to the acceleration parameter $q_{0}\approx-0.68$
can be found in \citep{Luo2015Dark,Luo:2015pca,Luo:2019iby,Luo:2021zpi}. 

If we include matter into the gravity theory, consider the entangled
system in $\mathcal{H}_{\psi}\otimes\mathcal{H}_{X}$ between the
under-studied quantum body and the quantum reference frame fields
system. Without loss of generality, we could take a scalar field $\psi$
as the under-studied system, which shares the base space with the
frame fields, the total action of the two entangled systems is a direct
sum of each system
\begin{equation}
S[\psi,X]=\int d^{d}x\left[\frac{1}{2}\frac{\partial\psi}{\partial x_{a}}\frac{\partial\psi}{\partial x_{a}}-V(\psi)+\frac{1}{2}\lambda g_{\mu\nu}\frac{\partial X^{\mu}}{\partial x_{a}}\frac{\partial X^{\nu}}{\partial x_{a}}\right],
\end{equation}
where $V(\psi)$ is some potential of the $\psi$ fields. Since both
$\psi$ field and the frame fields $X$ share the same base space
(fiducial lab), here they are formulated on the usual lab spacetime
$x$. If we interpret the frame fields as the physical spacetime coordinates,
the coordinates or reference frames of $\psi$ field must be transformed
from $x$ to $X$. At the semi-classical level, or 1st moment approximation
when the fluctuation of $X$ can be ignored, it is simply a coordinates
transformation
\begin{align}
S[\psi,X]\overset{(1)}{\approx}S[\psi(X)] & =\int d^{4}X\sqrt{|\det g^{(1)}|}\left[\frac{1}{4}\left\langle g_{\mu\nu}^{(1)}\frac{\partial X^{\mu}}{\partial x_{a}}\frac{\partial X^{\nu}}{\partial x_{a}}\right\rangle \left(\frac{1}{2}g^{(1)\mu\nu}\frac{\delta\psi}{\delta X^{\mu}}\frac{\delta\psi}{\delta X^{\nu}}+2\lambda\right)-V(\psi)\right]\nonumber \\
 & =\int d^{4}X\sqrt{|\det g^{(1)}|}\left[\frac{1}{2}g^{(1)\mu\nu}\frac{\delta\psi}{\delta X^{\mu}}\frac{\delta\psi}{\delta X^{\nu}}-V(\psi)+2\lambda\right],\label{eq:eff-gravity-(1)}
\end{align}
in which $\overset{(1)}{\approx}$ stands for the 1st moment approximation,
and $\frac{1}{4}\left\langle g_{\mu\nu}^{(1)}\frac{\partial X^{\mu}}{\partial x_{a}}\frac{\partial X^{\nu}}{\partial x_{a}}\right\rangle =\frac{1}{4}\left\langle g_{\mu\nu}^{(1)}g^{(1)\mu\nu}\right\rangle =\frac{1}{4}D=1$
has been used. It is easy to see, at the semi-classical level, i.e.
only consider the 1st moment of $X$ while 2nd moment fluctuations
are ignored, the (classical) coordinates transformation reproduces
the scalar field action in general coordinates $X$ up to a constant
$2\lambda$, and the derivative $\frac{\partial}{\partial x_{a}}$
is replaced by the functional derivative $\frac{\delta}{\delta X^{\mu}}$.
$\sqrt{|\det g^{(1)}|}$ is the Jacobian determinant of the coordinate
transformation, note that the determinant requires the coordinates
transformation matrix a square matrix, so at semi-classical level
$d$ must be very close to $D=4$, which is not necessarily true beyond
the semi-classical level, when the 2nd moment quantum fluctuations
are important. For instance, since $d$ is a parameter but an observable
in the theory, it could even not necessary be an integer but effectively
fractal at the quantum level, and we have chosen $d=4-\epsilon$.

When the gravity and quantum spacetime frame fields are normalized
by the Ricci flow, $2\lambda$ term in eq.(\ref{eq:eff-gravity-(1)})
is normalized by 2nd moment fluctuation, by using eq.(\ref{eq:eff-EH})
and eq.(\ref{eq:eff-R}), a matter-coupled-gravity is emerged from
the Ricci flow
\begin{align}
S[\psi,X]\overset{(2)}{\approx} & \int d^{4}X\sqrt{|\det g_{k}|}\left[\frac{1}{2}g^{\mu\nu}\frac{\delta\psi}{\delta X^{\mu}}\frac{\delta\psi}{\delta X^{\nu}}-V(\psi)+2\lambda-\frac{R_{0}}{64\pi^{2}}k^{2}+\lambda\nu\right]\nonumber \\
= & \int d^{4}X\sqrt{|\det g_{k}|}\left[\frac{1}{2}g^{\mu\nu}\frac{\delta\psi}{\delta X^{\mu}}\frac{\delta\psi}{\delta X^{\nu}}-V(\psi)+\frac{R_{k}}{16\pi G}+\lambda\nu\right]
\end{align}

\section{Conformal Stability of Local Quantum Spacetime and the F-functional}

Note that, defined by the variation of the effective action of gravity
(\ref{eq:relative-partition}) or the Shannon entropy (\ref{eq:shannon entropy}),
there is also a ``wrong sign'' in the F-functional for the spacetime
Ricci flow with $D=4$
\begin{equation}
\mathcal{F}(g,u)\equiv\frac{\partial N}{\partial\tau}=\int d^{D}Xu\left(R+|\nabla f|^{2}\right).
\end{equation}
It is formally similar with the Euclidean Einstein-Hilbert action
(\ref{eq:EH+conformal}) in which $d^{D}X$ plays the role of the
scale $k$ dependent measure $d^{4}x\sqrt{|\tilde{g}_{k}|}$ in the
action. The functional is a generalization of Perelman's F-functional
of a 3-space \citep{perelman2002entropy}, so it can be considered
for the 4-spacetime with Euclidean signature.

From the point of view of the F-functional, the conformal stability
of a spacetime configuration is determined by the sign of the lowest
eigenvalue of an operator $-4\Delta+R$ ($\Delta$ the Laplace-Beltrami
operator and $R$ the scalar curvature for the spacetime) associated
with the F-functional $\mathcal{F}(g,\phi)$ \citep{cao2004gaussian,Suneeta:2008is},
\begin{equation}
\Lambda(g)=\inf\left\{ \mathcal{F}(g,\phi)=\int d^{4}X\left(R\phi^{2}+4|\nabla\phi|^{2}\right),\;\textrm{with}\;\lambda\int\phi^{2}d^{4}X=1\right\} 
\end{equation}
rather than naively depends on the sign of the local kinetic term.
Just as the stability of quantum hydrogen atom, the competition between
the kinetic term $|\nabla\phi|^{2}$ and the potential term $R\phi^{2}$
is important for the conformal stability.

More precisely, to probe the stability of a local compact (closed
and bounded) spacetime region centered at spacetime point $X$, one
could choose $\phi^{2}(X)=u(X)=e^{-f(X)}$ an approximate delta function
centered at $X$ to calculate $\Lambda(g(X))$. From the backwards
heat equation (\ref{eq:u-t-eq}) and (\ref{eq:u-normalization}),
it is clear that if the eigenvalue $\Lambda(g(X))>0$, the local compact
4-volume $d^{4}X\sim u^{-1}$ of the region will decay and shrink
by the increasing of $u$ along the flow-time $t$ (but the physical-time)
into a shrinking soliton and tend to disappear (not completely collapse,
see later) or develop local neckpinch and tend to thorns in finite
flow-time $t$ (finite scale) during the Ricci flow, so the local
compact region around $X$ is linearly unstable. The positive eigenvalue
$\Lambda$ is similar with the negative unbounded classical action
$S_{E}$, leading to a conformally unstable region, and the function
$\phi$ plays a similar role of the conformal factor $\Omega$, not
only they both have similar ``wrong sign'' in their kinetic terms,
but also $\phi=\sqrt{u}$ represents the longitudinal and trace part
of the degrees of freedom of metric. Actually $\phi=\sqrt{u}$ is
indeed a conformal factor of the gravity up to a constant multiple,
a positive sign of $\Lambda$ will indeed induce a conformal instability
of gravity at least in a local compact region. Furthermore, it is
also found \citep{Luo:2021wdh} that such local unstable shrinking
region modeled by a shrinking Ricci soliton reproduces a spatial inflationary
region at a local physical-time. If $\Lambda(g(X))\le0$, the spacetime
compact region expands or keeps its volume, and hence it is stable
up to a rescaling, so it is equivalent to the positive mass theorem
in that region. The flat spacetime with $R=0$ is an example of $\Lambda=0$
which is conformally stable. 

Certainly, the similarity between the F-functional and the Einstein-Hilbert
action is formal but exact. From the previous effective action of
gravity (\ref{eq:partition with nu}) with long distance approximation
(\ref{eq:Ntild=00003Dtau*Ftild}), we can see that (up to a normalization
factor) the effective action given in terms of the Shannon entropy
$N$ differs from the F-functional by a $\tau$ parameter. $\mathcal{F}=\frac{\partial N}{\partial\tau}=0$
plays the role of a fixed point action (corresponding to the steady
soliton configuration), and $\tilde{\mathcal{F}}=\frac{\partial\tilde{N}}{\partial\tau}=0$
the fixed point action of the shrinking soliton, while the relative
Shannon entropy $\tilde{N}=N-N_{*}$ plays the role of the scale dependent
effective action. In this sense, to investigate the final destiny
of the unstable region (where $\Lambda>0$) and the stability of the
whole spacetime we need to study the exact action $\tilde{N}$ and
the related entropy functional. 

\section{Local Non-Collapsing of Quantum Spacetime and the W-functional}

What is the final destiny of the unstable region, does the local unstable
compact region completely collapse into nothing? Or even worse, does
the instability finally leads to a disastrous collapse of the whole
spacetime? A no-go answer to the second question seems guaranteed
by the positive mass theorem \citep{Schoen:1979zz,Edward1981A} at
the classical level, but what is the case at the quantum level?

The monotonicity of the F-functional $\frac{\partial\mathcal{F}}{\partial t}\ge0$
(see eq.(\ref{eq:F-monotonic})) claims that $\Lambda(g)$ is nondecreasing
along the flow-time $t$ during any Ricci flow without any curvature
condition, and bounded above by $\mathcal{F}\le\frac{D}{2\lambda(t_{*}-t)}=\frac{D}{2\lambda\tau}$
(followed from (\ref{eq:Ntilde_monontonic})), where $D$ is the dimension
of spacetime, $t$ the flow-time, $t_{*}$ is certain finite singular
flow-time when the local curvature diverges. The equal sign is saturated
if and only if the local spacetime region flows and finally becomes
to a gradient shrinking Ricci soliton configuration eq.(\ref{eq:shrinker}).

However, the right hand side of $\mathcal{F}\le\frac{D}{2\lambda(t_{*}-t)}$
also diverges when $t\rightarrow t_{*}$, the bound still does not
answer the question whether the local shrinking Ricci soliton region
finally local collapse as $t\rightarrow t_{*}$. More precisely, to
define the notion ``local collapse'' of a compact local spacetime
region, we consider if there exists a sequence ($i=1,2,3,...$) of
flow-time $t_{i}\rightarrow t_{*}$ and radii $r_{i}\in(0,\infty)$
of the local compact space and time region with $r_{i}^{2}/t_{i}$
uniformly bounded from above, and the Riemannian curvature of the
local region is bounded comparable to the radius $\left|Rm\left(g(t_{i})\right)\right|\le r_{i}^{-2}$
in the compact spacetime ``ball'' $B_{g(t_{i})}(r_{i})$, here the
volume of the spacetime ``ball'' in the local compact region 
\begin{equation}
V\left(B_{g(t_{i})}(r_{i})\right)\equiv\int_{B(r_{i})}d^{4}X_{t_{i}}
\end{equation}
shrinks to zero in the limit of the sequence $\lim_{i\rightarrow\infty}\frac{V\left(B_{g(t_{i})}(r_{i})\right)}{r_{i}^{4}}=0$.
If the situation actually happens, the local compact spacetime region
is said to be local collapse. 

To probe whether the local compact region is local collapse, we need
a dimensionless and scale invariant version of the F-functional. To
achieve the scale invariance, Perelman includes an explicit insertion
of the scale parameter $\tau=t_{*}-t$ to the F-functional and defines
the W-functional. From the anomaly by using eq.(\ref{eq:W=00003DtauF+N}),
the Legendre transformation of the relative Shannon entropy $\tilde{N}$
w.r.t. $\tau^{-1}$ gives
\begin{equation}
\mathcal{W}\equiv\tau\frac{\partial\tilde{N}}{\partial\tau}+\tilde{N}=\int d^{D}Xu\left[\tau\left(R+|\nabla f|^{2}\right)+f-D\right]
\end{equation}
formally the same with Perelman's W-functional with $D=4$ the dimension
of the spacetime, and $u$ defined by (\ref{eq:u}) is the positive
defined manifolds density allowing one to localize to the concern
compact region, $\tau=t_{*}-t$ telling us at what distance scale
to localize (i.e. $\sqrt{\tau}$). The W-functional will be essential
for understanding the critical structure near a shrinking spacetime
region. First, it is invariant under simultaneous rescaling of $\tau$
and $g$. Second, it is non-decreasing along the flow-time during
any Ricci flow $\frac{\partial\mathcal{W}}{\partial t}\ge0$ (see
eq.(\ref{eq:W_monotonic})) without any assumption on curvature, for
this reason, it is also often called the W-entropy. 

One can also use the $u$ density to probe the local compact region
of $g(t)$ where one concerns. For example, the collapse or non-collapse
of the region near a point $X$ can be detected from the value of
the W-functional. If one chooses $u(X,t)$ an approximate delta function
centered at $X$ at flow-time $t$, then the more collapse of the
region, the more negative the value of $\mathcal{W}\left[g(X),u(X),t\right]$.
However, by the monotonicity eq.(\ref{eq:W_monotonic}) of the W-functional
during the Ricci flow, since the W-functional of a certain initial
metric of the region is bounded from below at certain initial scales
(bounded from above and below), after a certain amount of flow-time,
the W-functional must also be bounded from below at all bounded scales,
so the local collapsing of the compact spacetime region corresponding
to the arbitrary negative value of the W-functional must be ruled
out by the monotonicity of it. 

This leads to the (4-spacetime generalization of) local non-collapsing
theorem of Perelman \citep{perelman2002entropy}, which states that
if a local compact spacetime region around point $X$ is unstable
and hence tends to shrink, as one approaches the singular finite flow-time
$t_{*}$ when the local curvature diverges, collapsing of the spacetime
region cannot actually occur at the scales $r\sim O(\sqrt{t_{*}-t})$,
the volume of a unit ball in the compact space and time region or
essentially the volume ratio is bounded from below
\begin{equation}
\frac{V\left(B_{g(t)}(X,r)\right)}{r^{D}}\ge\kappa(g,D)>0.
\end{equation}

The local non-collapsing theorem has important physical consequences.
Since the inequality $\frac{\partial\mathcal{W}}{\partial t}\ge0$
saturates when the spacetime region is a gradient shrinking Ricci
soliton eq.(\ref{eq:shrinker}). Although the volume of the shrinking
soliton is shrinking due to the conformal instability, the scale invariant
$\mathcal{W}$ entropy has already maximized to be a finite constant
value, the local shape of the shrinking soliton does not change, the
information of its shape or topology is preserved rather than lost
(when $\mathcal{W}$ entropy is arbitrarily negative and hence collapse,
then the information is completely lost). The information of its size
or volume is not preserved (relative to an observer and depends on
the definition of ruler and clock of the observer), but its structure
encoded by its shape and topology can always be ``zoomed in'' by
the observer if it does not completely collapse into nothing. More
physical speaking, the conformal instability may shrink a local compact
spacetime region but will not collapse it into nothing, otherwise
the scale-invariant residue information of its local topology will
be lost during the Ricci flow. As a consequence, local conformal instability
may happen in some places where $\Lambda(g(X))>0$ for a given general
initial spacetime manifolds, and the local Lagrangian is not necessarily
positive defined, but the total effective action (\ref{eq:relative-partition})
$S_{eff}(M^{D})=-\log Z=\frac{D}{2}-\lambda\tilde{N}(M^{D})$ should
be positive defined and bounded from below, where the effective action
$\frac{D}{2}-\lambda\tilde{N}(M^{D})$ is given by QSRF corresponding
to Perelman's partition function \citep{perelman2002entropy} in the
Ricci flow. It is easy to verify that it is indeed the case, because
the Relative Shannon entropy $\tilde{N}(M^{D})$ is bounded above
$\tilde{N}(M^{D})\le0$ from eq.(\ref{eq:perelman-partition}) as
is shown in Section-II.

The positive and boundedness of the total effective action $S_{eff}(M^{D})$
ensures the stability of the whole spacetime, which can be seen as
a quantum generalization of the classical positive mass theorem in
certain sense. In the geometric point of view, starting from an arbitrary
initial spacetime manifolds, the Ricci flow as a RG flow gradually
deforms and smooths out local irregularities on it, after certain
proper treatment of the local neckpinch (``surgeries'' by hand \citep{perelman2003ricci}
or by internal mechanism of the full RG flow) the spacetime will finally
flow to a stable and non-collapse spacetime up to a rescaling. 

\section{Discussions and Conclusions}

In this paper, the framework quantum spacetime reference frame (QSRF)
is reviewed. Differing from other routes, the framework provides us
an alternative way to generalize a 3-space Ricci flow to a Riemannian
or Lorentzian 4-spacetime version and has clear physical meaning.
Under the physical interpretation of QSRF, the volume form and density
$u$ defined by eq.(\ref{eq:volume form}) in the framework ensure
the monotonic functionals induced from the anomaly of QSRF formally
the same with Perelman's standard functionals in a Riemannian or Lorentzian
spacetime. The Ricci flow is a gradient flow of the monotonic functionals,
the variational structure makes the functionals suitable for studying
the stability problem of spacetime.

The conformal instability problem of quantum spacetime is studied
in the framework of QSRF and induced spacetime Ricci flow. Instead
of naively observing a ``wrong sign'' in front of the kinetic term
of the conformal factor in the path integral of general relativity,
the conformal stability of quantum spacetime and gravitation actually
depend on the sign of the lowest eigenvalue $\Lambda(g(X))$ of the
operator $-4\Delta+R$, associated with the F-functional around a
local compact (closed and bounded) spacetime region. If $\Lambda(g(X))>0$
the local compact region is conformally unstable and will tend to
shrink its volume (along the flow-time $t$ but the physical-time),
if $\Lambda(g(X))\le0$ the local compact region is conformally stable
up to a trivial rescaling. Thus a given general spacetime will possibly
develop local volume shrinking and local curvature pinching in some
places. Although conformal instability may happen in some places of
a given general spacetime, the instability will not cause the structure
of the local region collapse into nothing, a finite residue information
taken by the scale-invariant W-functional of that region is preserved.
The residue information of the shrinking local region will not be
lost, otherwise the W-entropy will become infinitely negative which
is ruled out by the monotonicity of it.

Different from the possible approaches to the conformal instability
of the Euclidean general relativity in literature, which try to locally
flip the ``wrong sign'' into a right one in the Lagrangian, the
framework of QSRF and induced Ricci flow allow the local Lagrangian
in the functional integral being negative, leading to conformally
unstable in certain places of the spacetime, but the total effective
action is proved positive defined and bounded below, so the whole
spacetime is stable, which generalizes the classical positive mass
theorem of spacetime to the quantum level. 

The presence of the conformal instability of a quantum spacetime may
have important physical consequence, for instance, an inflationary
universe at early epoch may be driven by the mechanism of the conformal
instability \citep{Luo:2021wdh}, in other words, the spatial inflationary
region near the local physical-time origin may be modeled by a gradient
shrinking Ricci soliton configuration, leading to a possible quantum
theory of early universe different from the textbook inflation models
driven by inflaton. The instability may also lead to the formation
of black hole as a temporal static shrinking Ricci soliton \citep{Luo:2022statistics}.
The conformal instability tends to shrink the local spacetime volume
but keep it non-collapsing, it implies a possible way to test the
work is to probe (directly or indirectly) the characteristic spectrum
of quantum fluctuations produced from the shrinking soliton-like configurations
such as the early universe and black hole, which are worth further
studying.

We also notice limitations of the framework. The basic limitation
is that it is not clear if the monotonic functionals of Perelman's
type are mathematically well-defined for the Lorentzian 4-spacetime,
a rigorous mathematical foundation is still under development. In
this paper, based on the proposed physical foundation, i.e. the QSRF
approach, the generalization of Ricci flow and its monotonic functionals
to 4-spacetime by using the relativistic frame fields theory is formal
and direct. In general, ghost states and/or horizon effects may arise
in a Lorentzian or indefinite-metric quantum theories, it is not clear
if it is well-defined or has richer consequences from the quantum
Lorentzian spacetime than the standard Riemannian version, or just
as is shown in the paper that the two versions are formally the same
due to the fact that they share common definitions of the positive-defined
volume forms $\sqrt{|g|}$ and the manifolds density $u$.
\begin{acknowledgments}
This work was supported in part by the National Science Foundation
of China (NSFC) under Grant No.11205149, and the Scientific Research
Foundation of Jiangsu University for Young Scholars under Grant No.15JDG153.
\end{acknowledgments}

\bibliographystyle{unsrt}

\end{document}